\documentclass[a4paper]{jpconf}
\usepackage{graphicx}
\begin{document}
\title{Status of reaction theory for studying rare isotopes}

\author{F.M. Nunes$^{1,2}$ and N.J. Upadhyay$^1$}

\address{$^1$National Superconducting Cyclotron Laboratory, Michigan State University, East Lansing, MI 48824, USA}
\address{$^2$Department of Physics and Astronomy, Michigan State University, East Lansing, MI 48824, USA}

\ead{nunes@nscl.msu.edu}

\begin{abstract}
Reactions are an important tool to study nuclear structure and for extracting  reactions relevant for astrophysics. In this paper we focus on deuteron induced reactions which can provide information on neutron shell evolution as well as neutron capture cross sections. We review recent work on the systematic comparison of the continuum discretized coupled channel method, the adiabatic wave approximation and the Faddeev momentum-space approach. We also explore other aspects of the reaction mechanism and discuss in detail difficulties encountered in the calculations.
\end{abstract}

\section{Introduction}
\vspace{0.5cm}

Nuclear reactions are one of the most powerful tools to explore the properties of isotopes over the whole
nuclear landscape. In addition, from nuclear reaction measurements one can extract capture rates relevant for astrophysics. For these two reasons, it is pivotal to develop the most reliable reaction theory and determine the accuracy with which nuclear reaction predictions can be made.

Rare isotopes have been critical to our understanding of the stability of matter, providing information on the various components of the effective interaction that binds neutrons and protons together. As one explores nuclear properties along isotopic chains, one can establish the isospin dependence of the interaction. As one explores nuclear properties at the limits of stability, one gains insight on the relevance of the continuum.

Rare isotopes also play a very important role in astrophysics. Today it is understood that a good fraction of the heavy elements created in the universe, are produced in the r-process. In this process, a rapid sequence of neutron capture takes us many nucleons away from stability, generating a path through very neutron-rich isotopes. Inputs to astrophysical simulations of the r-process require masses, lifetimes and neutron-capture rates for these exotic nuclei. Since neutron capture on an exotic nucleus is presently not experimentally feasible, alternate methods using nuclear reactions have been proposed.
Of primary interest is the use of (d,p) reactions for this purpose.

This is a particularly exciting time for nuclear physics.
New rare isotope facilities have either just come online (e.g. RIBF at RIKEN) or  are being constructed
(e.g. FRIB at MSU). These new facilities, not only have a wider reach into unknown regions, but also have increased rates which will allow for many more, and more detailed, reaction measurements. Since the analysis of the reaction measurements require reaction theory, we underline the importance of having control over the ambiguities in the reaction calculations.

Deuteron induced reactions, being one of the simplest probes, enable inspection of both nuclear structure and nuclear astrophysics. Our work within the TORUS collaboration \cite{torus} has focused on benchmarking models for (d,p) reactions and advancing the theory for (d,p) to a new level, with reduced approximations and wider applicability.

Solving the many-body reaction problem A(d,p)B is not feasible whenever the large number of nucleons to be considered is larger than $A \approx 10$. The additional difficulty of the reaction many-body problem, as compared to the bound state problem, arises due to not only the behavior of the scattering wave-function at large distances but also due to the extreme sensitivity of reaction observables to thresholds and the correct asymptotic behavior of bound states in the subsystems. In most cases, it is thus necessary to reduce the many-body problem to a few-body problem, retaining only the relevant degrees of freedom. In particular, for A(d,p)B, this would imply a reduction to a three-body problem $n+p+A$. With this reduction, optical potentials, with their inherent ambiguities, are  added to our theory. In a three-body model for A(d,p)B, only nucleon optical potentials are need (apart from the NN interaction). These can be rather well constrained by nucleon elastic scattering and the associated uncertainty can be well estimated. The three-body formulation for A(d,p)B also contains many-body overlap functions which derive from the underlying many-body structure of the nuclei involved, A and B, after integrating unnecessary degrees of freedom. Evidently, (d,p) reactions are used to probe exactly this ingredient.

The three-body scattering problem can be solved exactly within the Faddeev momentum space integral formalism \cite{ags,deltuva09}  which explicitly includes breakup and transfer channels to all orders.
One of the most well established theories for direct nuclear reactions  is the
Continuum Discretized Coupled Channels (CDCC) method \cite{cdcc,scholar}. It includes
breakup to all orders by discretizing the projectile continuum into bins.
A systematic comparison of CDCC and Faddeev has been completed \cite{upadhyay12}. Another method, developed specifically for (d,p) transfer reactions, is the so-called adiabatic wave approximation (ADWA) \cite{tandy}. One can think of ADWA as a simplified CDCC calculation, where the radial behavior of all the continuum components are assumed to be similar to the ground state. It is thus expected to work best if the reaction does not excite higher energies in the projectile, i.e. low energy reactions. Comparison of this method and Faddeev has also been performed in a systematic manner \cite{nunes11}.
In this paper we summarize the results of these comparisons, explore other aspects of the reaction mechanism and discuss some challenges of the above mentioned calculations.

\section{Comparing methods for (d,p)}
\vspace{1cm}

The findings in \cite{upadhyay12} prove that CDCC is able to provide a good approximation to Faddeev
for elastic scattering. The explicit coupling to the elastic channel with the neutron-nucleus bound state as done in Faddeev calculations introduces small modifications to the elastic angular distributions mostly at backward angles.

The comparison of CDCC and Faddeev for transfer cross sections is consistent with the comparison of ADWA and Faddeev
presented in \cite{nunes11}.
We found CDCC and ADWA to be very good approximations of Faddeev for reactions around 10 MeV/u, while the agreement deteriorated with larger beam energies. At larger beam energies observables also become much more dependent on subtle differences in the interactions used in CDCC/ADWA versus Faddeev \cite{upadhyay12,nunes11}.  For this reason, the shortcomings of either CDCC or ADWA at larger beam energies are tamed by ambiguities in the choice of the energy at which the optical potentials are calculated.

It became clear from the comparisons in \cite{upadhyay12,nunes11} that, for projectiles with primarily a loosely bound s-wave ground state, CDCC did not improve the description of transfer when compared to the adiabatic model (ADWA). For these cases, ADWA is the preferred tool, given the fact that it is computationally much more efficient than CDCC. It was also found that at larger beam energies, CDCC did not show significant improvement over ADWA. This later point is still rather puzzling since CDCC does have a much better description of the n-p continuum as compared to ADWA.

The situation for breakup is very different.
Breakup observables predicted by CDCC are at its best for the higher beam energies explored in \cite{upadhyay12,nunes11}. At the low energies, CDCC results are far from the results from Faddeev. We note that to reduce the technical difficulties in the Faddeev approach, the Coulomb interaction was neglected when computing breakup observables (we will come back to this in 3.2).  In addition we underscore that exactly the
same Hamiltonian for CDCC and Faddeev was used and therefore, when comparing CDCC and Faddeev for breakup, there is no ambiguity in the choice of the interaction.

The  CDCC/FAGS comparisons in \cite{upadhyay12} demonstrate that finding agreement for the elastic for a given target and beam energy does not imply agreement in breakup or transfer. This results from sensitivity of the reaction amplitude to different parts of configuration space. Thus, only by looking at elastic, transfer and breakup simultaneously, the CDCC method can be thoroughly tested.

\begin{figure}[t!]
\begin{center}
\includegraphics[scale=0.45]{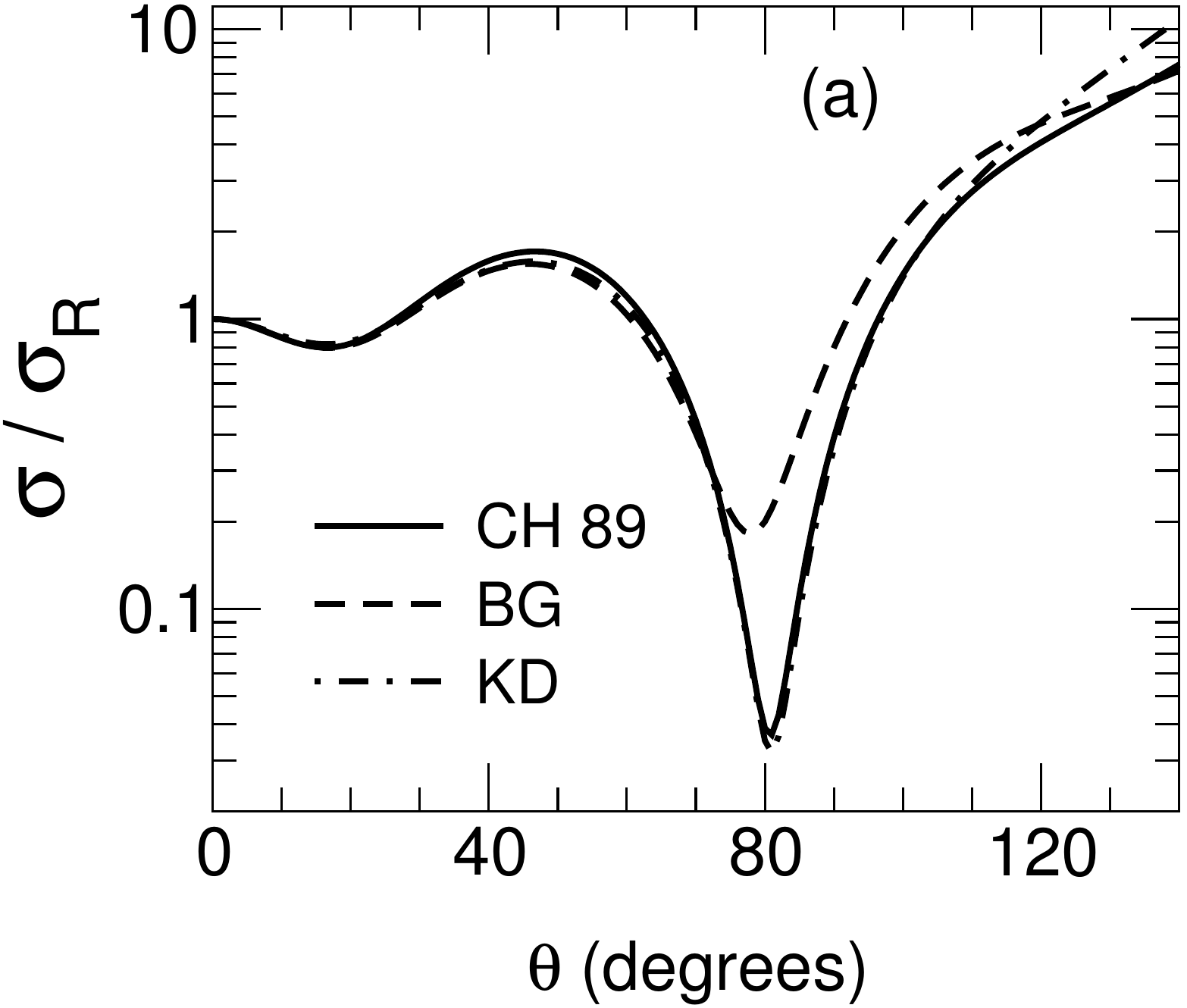}
\hspace{0.5cm}
\includegraphics[scale=0.45]{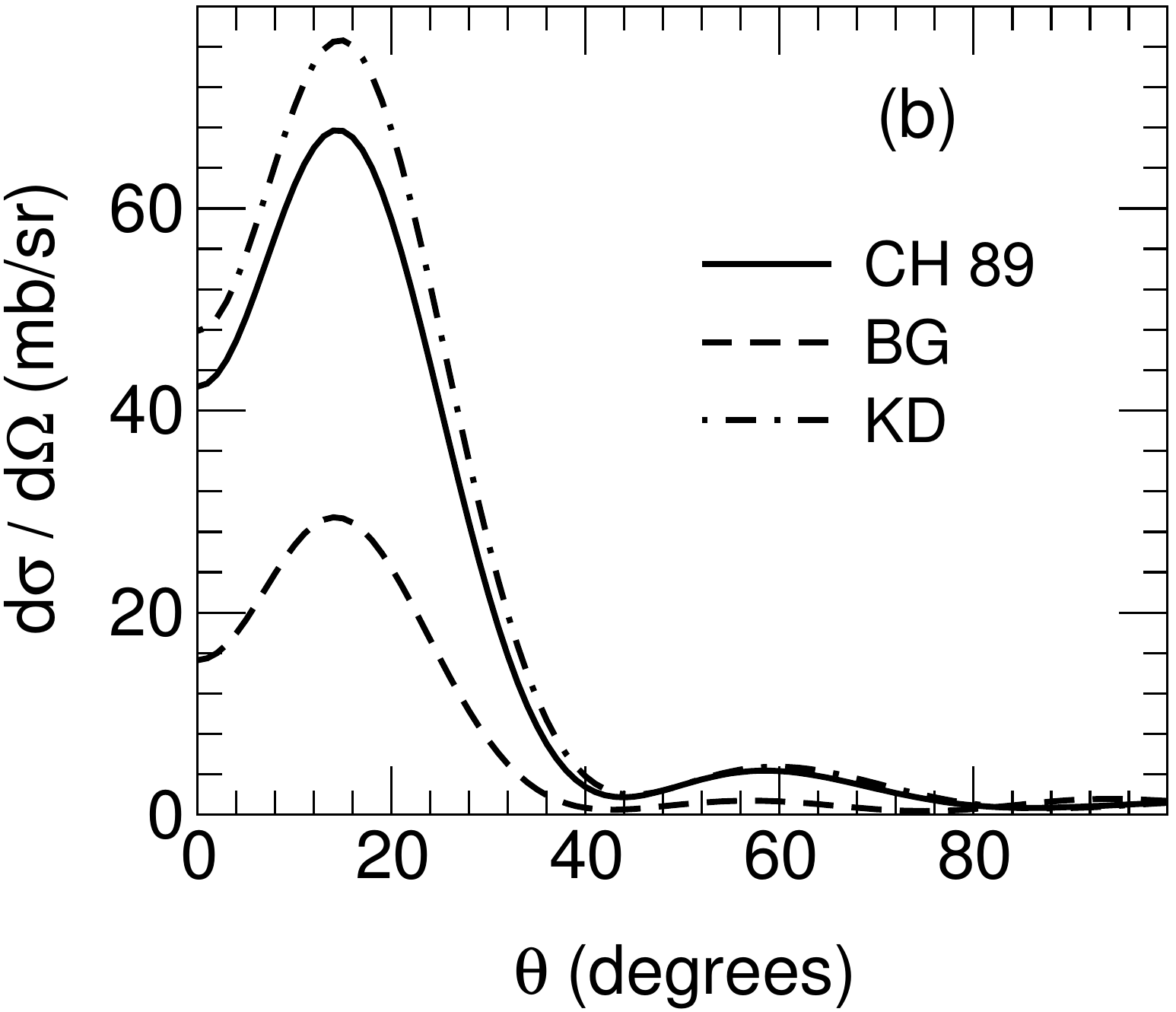}
\end{center}
\caption{\label{c12op}\small Optical model dependence: (a) elastic angular distribution for p$+^{12}$C at $E_{\rm p}$ =6 MeV and (b) transfer angular distribution for $^{12}$C(d,p)$^{13}$C at $E_{\rm d}$ = 12 MeV. Shown are results using Chapel-Hill89 \cite{ch89} (solid),  Bechetti and Greenlees \cite{bg} (dashed) and Konin and Delaroche \cite{kd} (dot-dashed). }
\end{figure}

\section{Discussion}
\vspace{0.5cm}

We further explore the results obtained in \cite{upadhyay12}, in regards to the dependence
of transfer observables on the interactions, dissecting the reaction mechanism for transfer reactions.
We also consider the difficulties of calculating breakup observables accurately, namely concerning
the treatment of Coulomb and the convergence of the cross sections with increasing model space.
The numerical inputs for the CDCC calculations performed in this section and the parameters used for the interactions are the same as those in \cite{upadhyay12}, unless otherwise stated. All calculations were performed using {\sc fresco} \cite{fresco}.

\begin{figure}[t!]
\begin{center}
\includegraphics[scale=0.45]{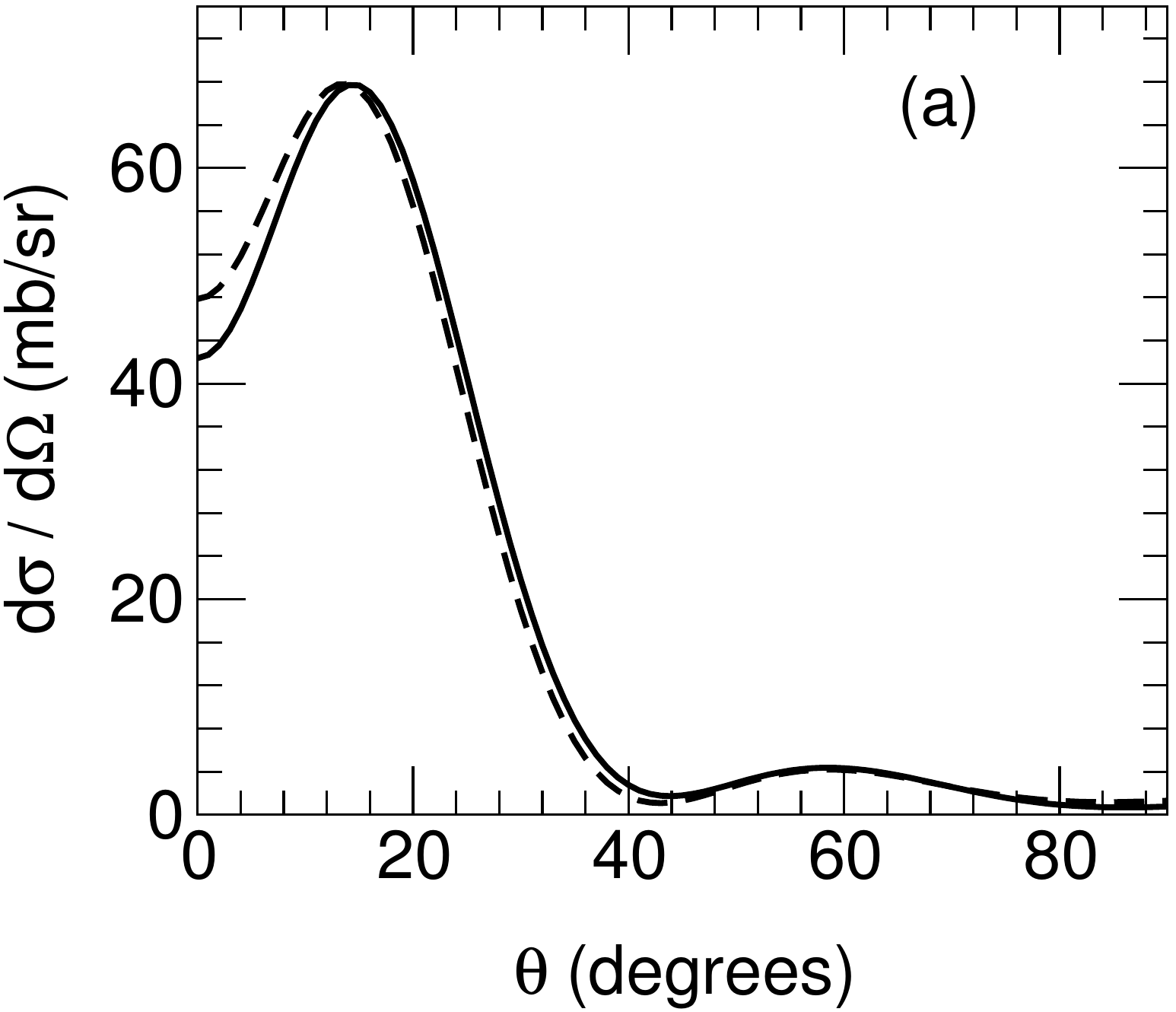}
\hspace{0.5cm}
\includegraphics[scale=0.45]{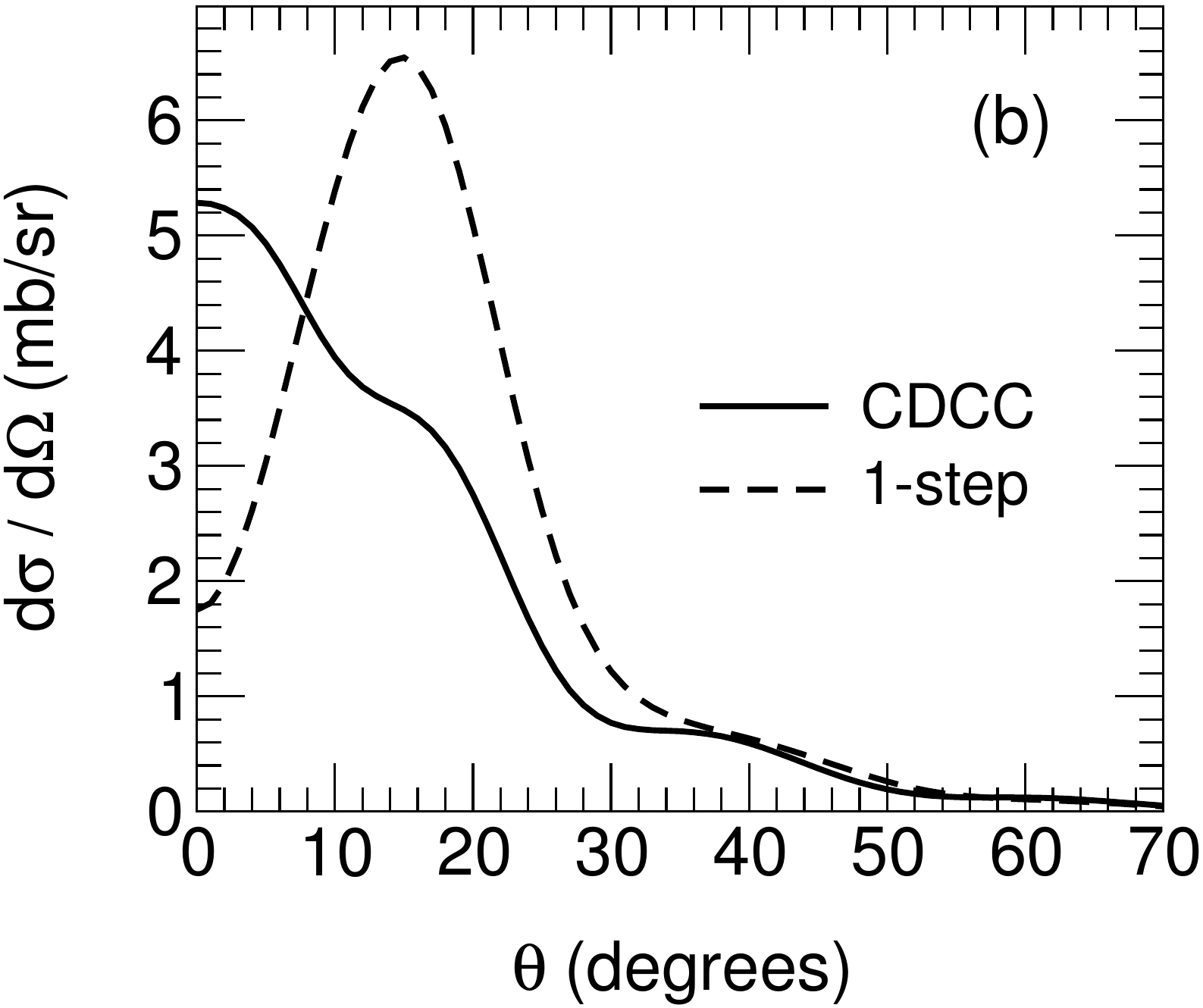}
\end{center}
\caption{\label{mechanism}\small Deuteron continuum effects on transfer cross sections: $^{12}$C(d,p)$^{13}$C at $E_{\rm d}$ = 12 MeV, (b) $^{48}$Ca(d,p)$^{49}$Ca at $E_{\rm d}$ = 56 MeV. Comparing full CDCC results (solid) with 1-step approximation (dashed). }
\end{figure}
\subsection{Looking at transfer in more detail}

As mentioned in the introduction, the reduction of the many-body problem to a few-body problem brings up the issue of optical potential ambiguities which need to be kept under control. As microscopic approaches improve, it will eventually become possible to extract optical potentials directly from the solutions of the many body problem (e.g. \cite{prl-thompson}). At present the most reliable way to constrain the optical potential is through elastic scattering. In Fig.\ref{c12op} we show the proton angular distribution for $^{12}$C at 6 MeV for three different optical potential parameterizations (fig. a) and the corresponding (d,p) transfer angular distribution as a function of the center of mass angle (fig. b). While CH89 \cite{ch89} and KD \cite{kd} provide similar proton angular distributions for angles smaller than $100 \deg$ (they have been shown to work well for these low energies  and low mass targets), larger differences are seen for BG \cite{bg}. When the interaction is constrained by elastic scattering, the ambiguity that carries over to transfer is small (the percentage difference at the peak between CH89 and KD in Fig.\ref{c12op}(b) is $\approx 10$\%). However, the huge difference in the normalization of the transfer cross section obtained when using BG as compared to CH89/KD emphasizes the well known fact that not constraining the optical potential can introduce large errors in the calculation. It is thus essential that a serious scientific program using (d,p) reactions as a probe for nuclear structure, include systematic measurements of nucleon elastic scattering
on the relevant isotopes and at the relevant energies.

When using the CDCC wavefunction to compute transfer amplitudes, it is possible to isolate the effect
coming from the direct transfer process and the multi-step process via breakup states in the $np$ system.
In Fig.\ref{mechanism} we compare the importance of breakup as an intermediate step in the transfer process, for (d,p) reactions on $^{12}$C at 12 MeV (left panel) and $^{48}$Ca at 56 MeV (right panel). While in the first case the effects are minor, for $^{48}$Ca(d,p) at 56 MeV, breakup has such a large effect on transfer that it changes completely the shape of the angular distributions. For this reason the traditional DWBA (distorted wave Born approximation) should only be used in cases where breakup has been proven to be insignificant otherwise the extracted information will not be meaningful.

\begin{figure}[t!]
\begin{center}
\includegraphics[scale=0.45]{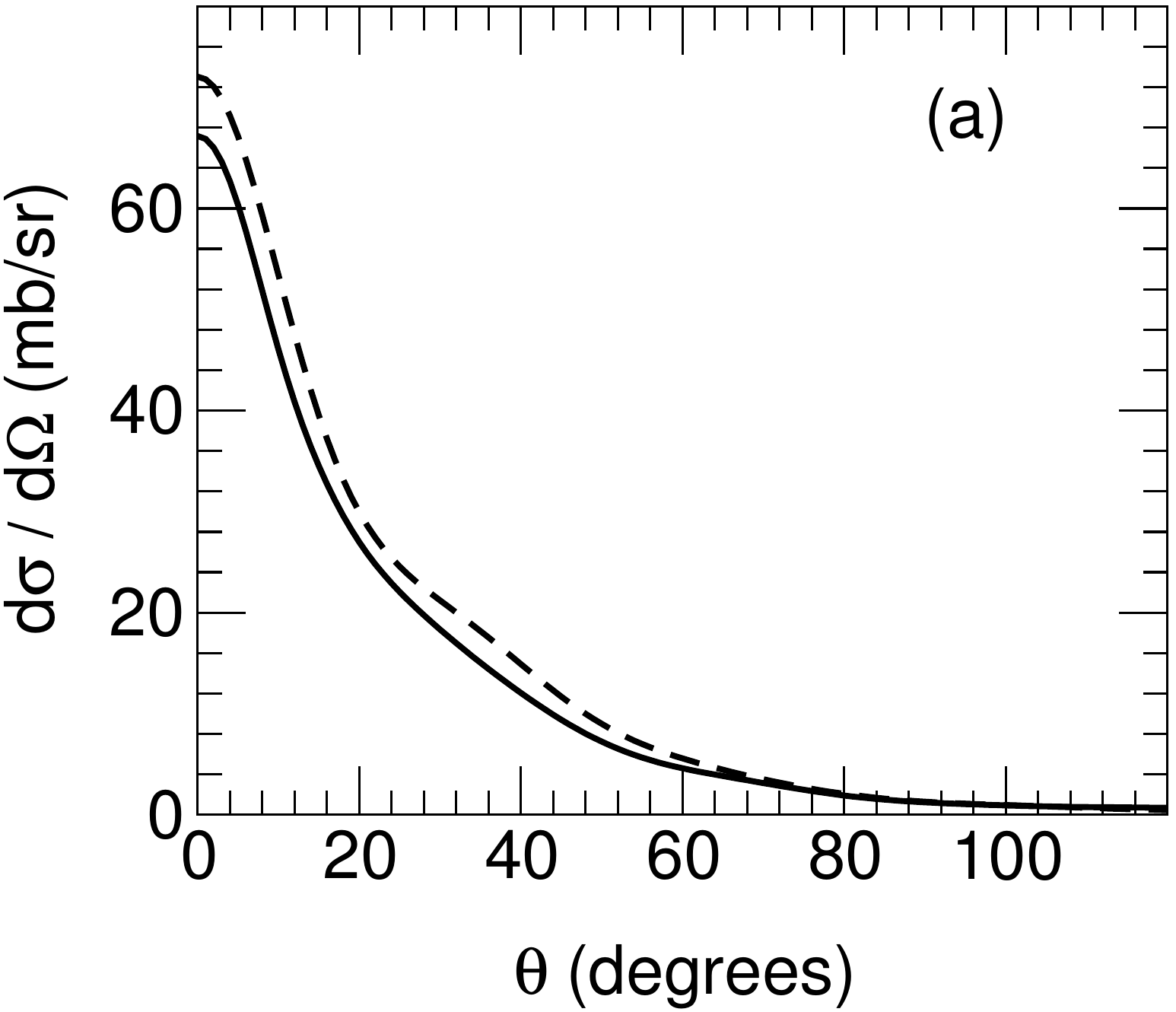}
\hspace{0.5cm}
\includegraphics[scale=0.45]{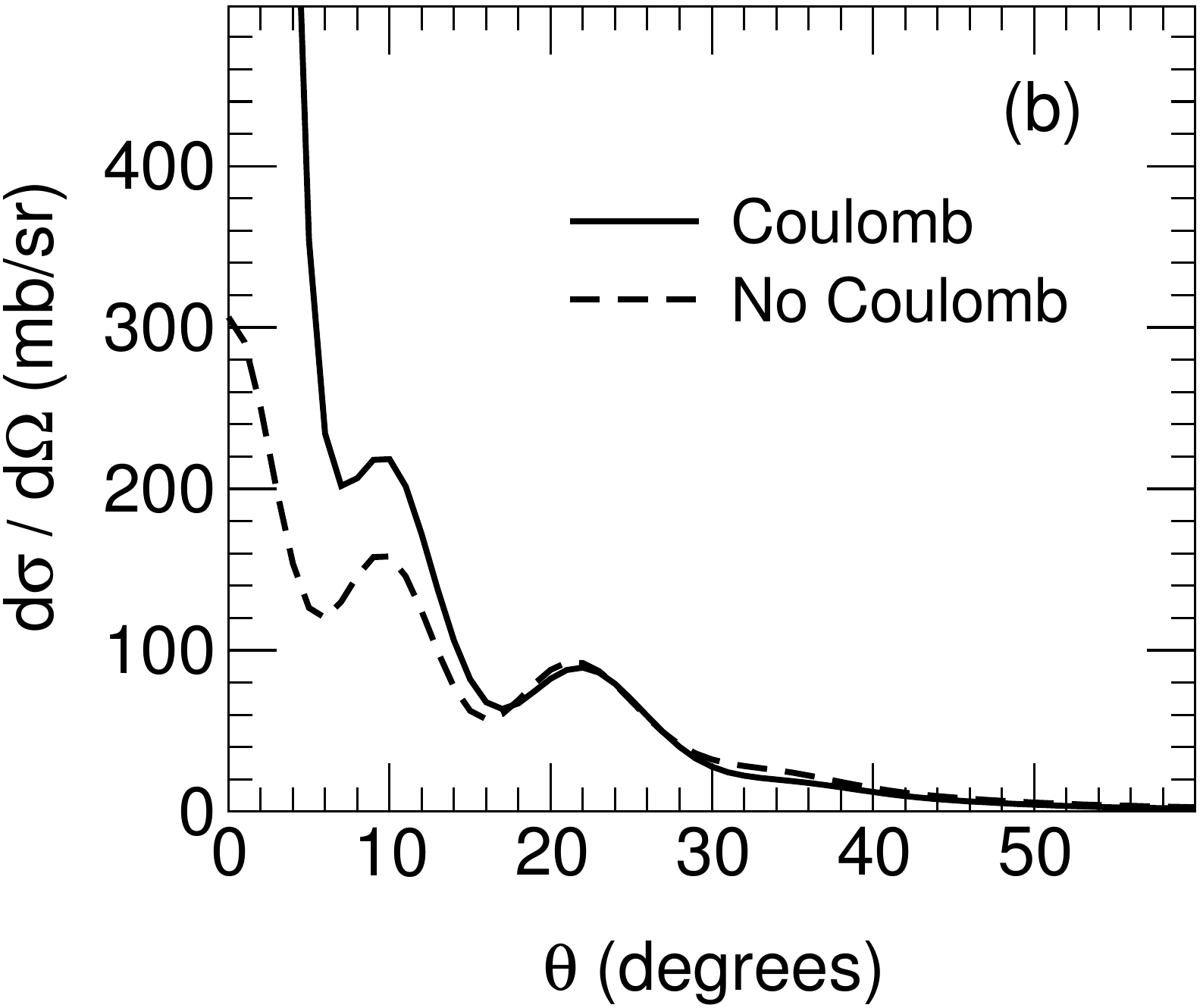}
\end{center}
\caption{\label{coulomb}\small Coulomb effects on the breakup cross sections: $^{12}$C(d,pn)$^{12}$C at $E_{\rm d}$ = 12 MeV, (b) $^{48}$Ca(d,pn)$^{48}$Ca at $E_{\rm d}$ = 56 MeV. Comparing the differential cross sections for the full CDCC results (solid) with those excluding Coulomb interactions (dashed). }
\end{figure}
\begin{figure}[t!]
\begin{center}
\includegraphics[scale=0.45]{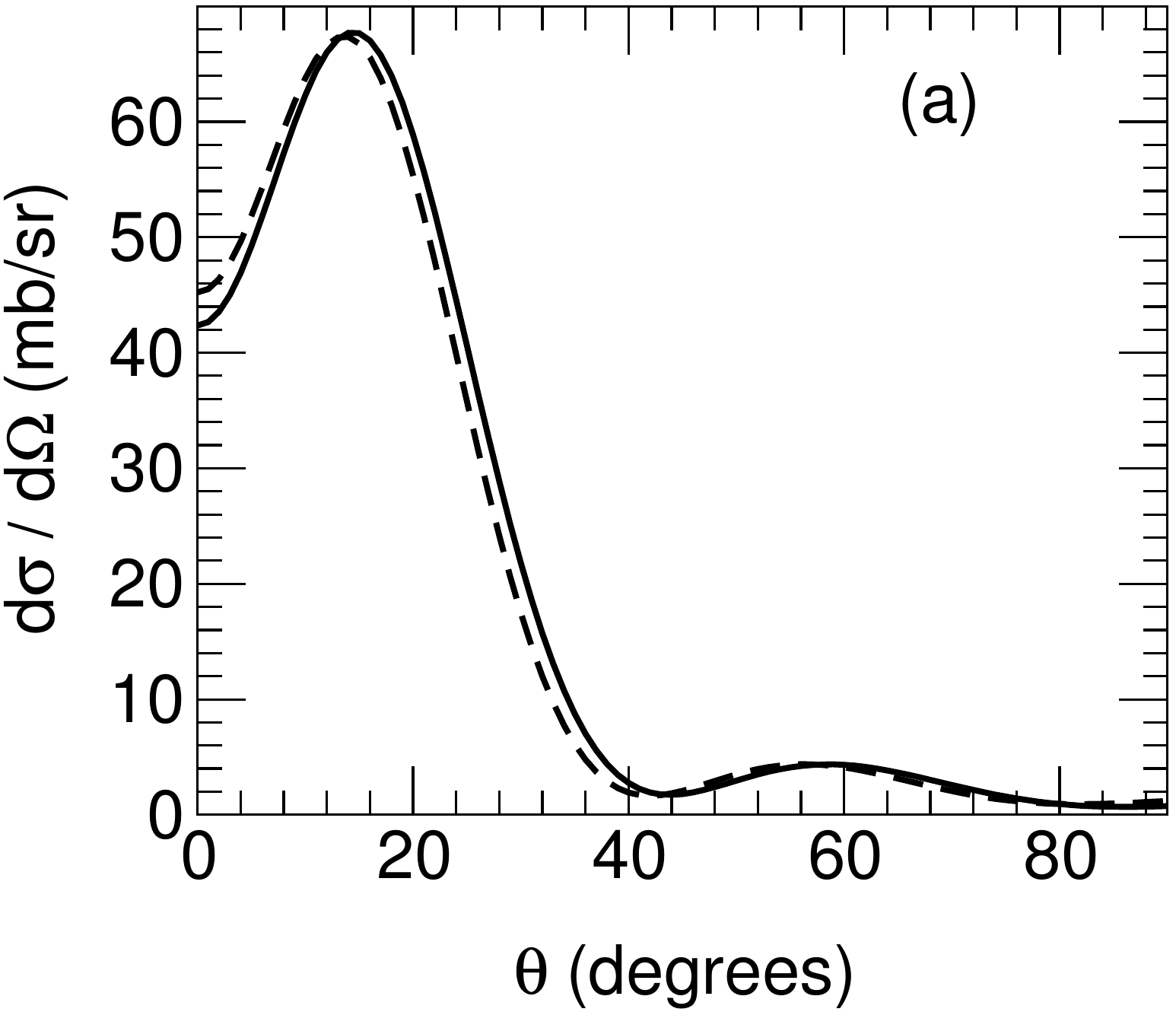}
\hspace{0.5cm}
\includegraphics[scale=0.45]{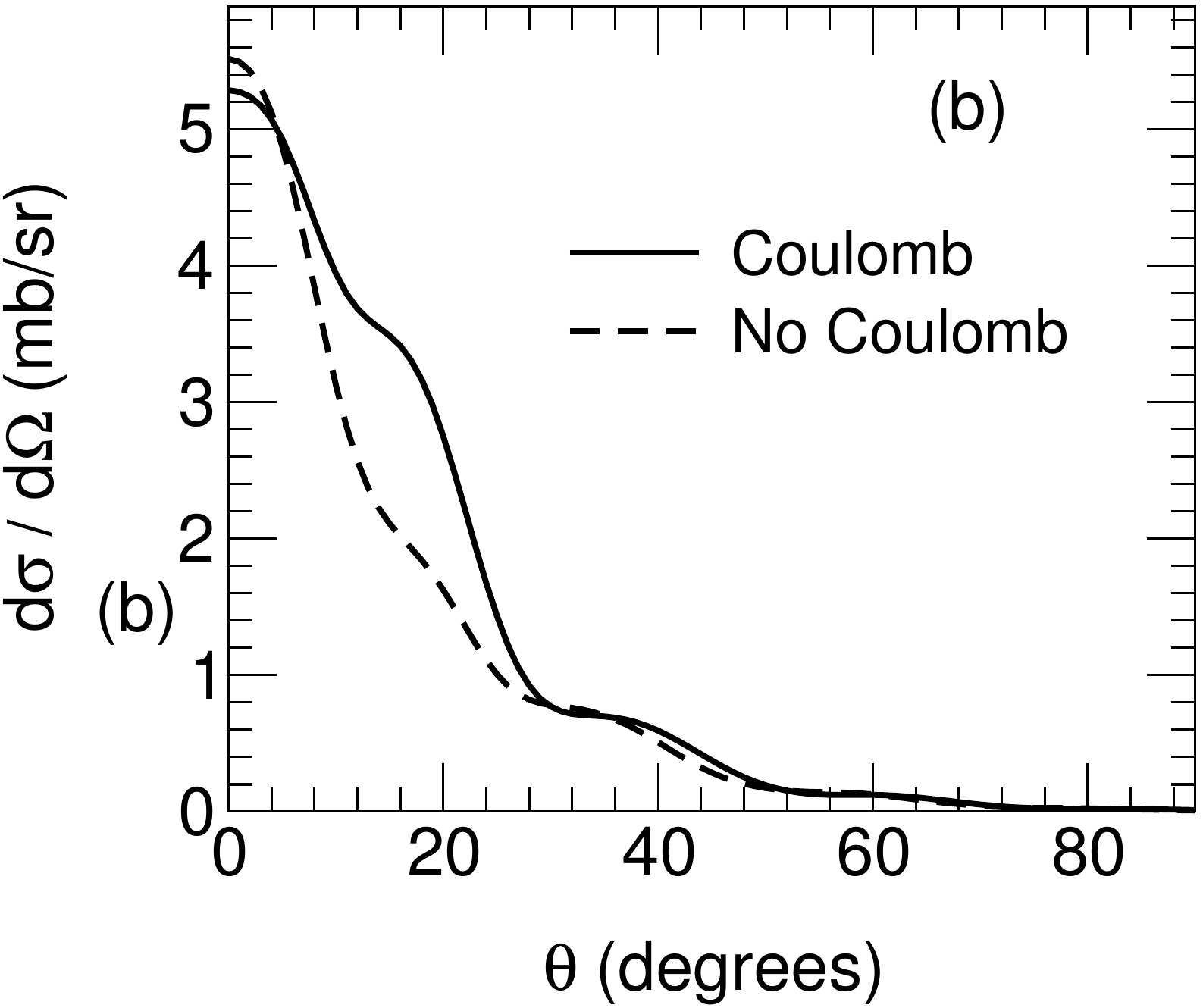}
\end{center}
\caption{\label{coulomb-tr}\small Coulomb effects on the transfer cross sections: $^{12}$C(d,p)$^{13}$C at $E_{\rm d}$ = 12 MeV, (b) $^{48}$Ca(d,p)$^{49}$Ca at $E_{\rm d}$ = 56 MeV. Comparing the differential cross sections for the full CDCC results (solid) with those excluding Coulomb interactions (dashed). }
\end{figure}

\subsection{Looking at breakup in more detail}

One important limitation to the Faddeev calculations performed in \cite{nunes11,upadhyay12} has to do with the Coulomb interaction. The momentum-space implementation relies on Coulomb  screening. While Coulomb screening works well for systems with small charge, as the Coulomb interaction becomes stronger, the screening radius necessary for convergence increases and with it the needed partial waves. Eventually the problem becomes numerically unstable. To avoid this known difficulty, breakup calculations presented in \cite{upadhyay12} excluded the Coulomb interaction.

The CDCC equations are solved in coordinate space and do not require screening. We thus use them to explore the relevance of the Coulomb interaction in deuteron induced reactions.  Fig.\ref{coulomb} shows the predictions for the breakup angular distributions as function of the center of mass angle of the outgoing $np$ system, for the full CDCC calculation including Coulomb (solid) and for a CDCC calculation where all charges are set to zero (dashed). Expectedly for light systems the Coulomb effects are small. However, for medium mass systems these effects become important, particularly at small angles, which captures the long-range nature of the Coulomb interaction.

In Fig.\ref{coulomb-tr} we show the corresponding plot for the transfer angular distributions.
Although effects on transfer observables are smaller than for breakup, Coulomb effects increase
with charge and these are expected to become even more important for heavy systems. With the techniques
of \cite{deltuva09}, reactions on heavy systems are not accessible. A new technique to handle Coulomb is being proposed within the scope of the TORUS collaboration \cite{akram12}.

\begin{figure}[t!]
\begin{center}
\includegraphics[scale=0.4]{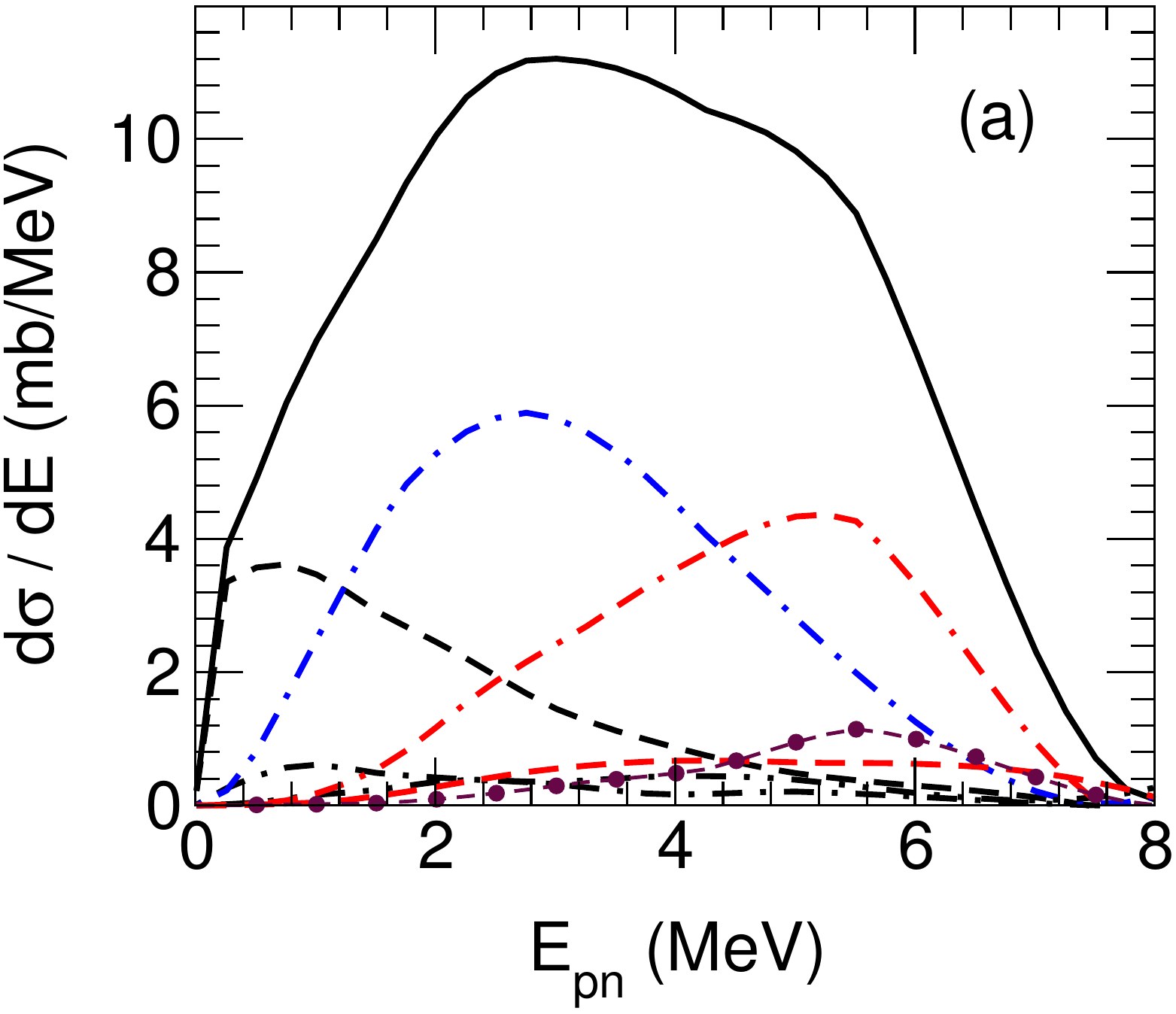}
\hspace{0.5cm}
\includegraphics[scale=0.4]{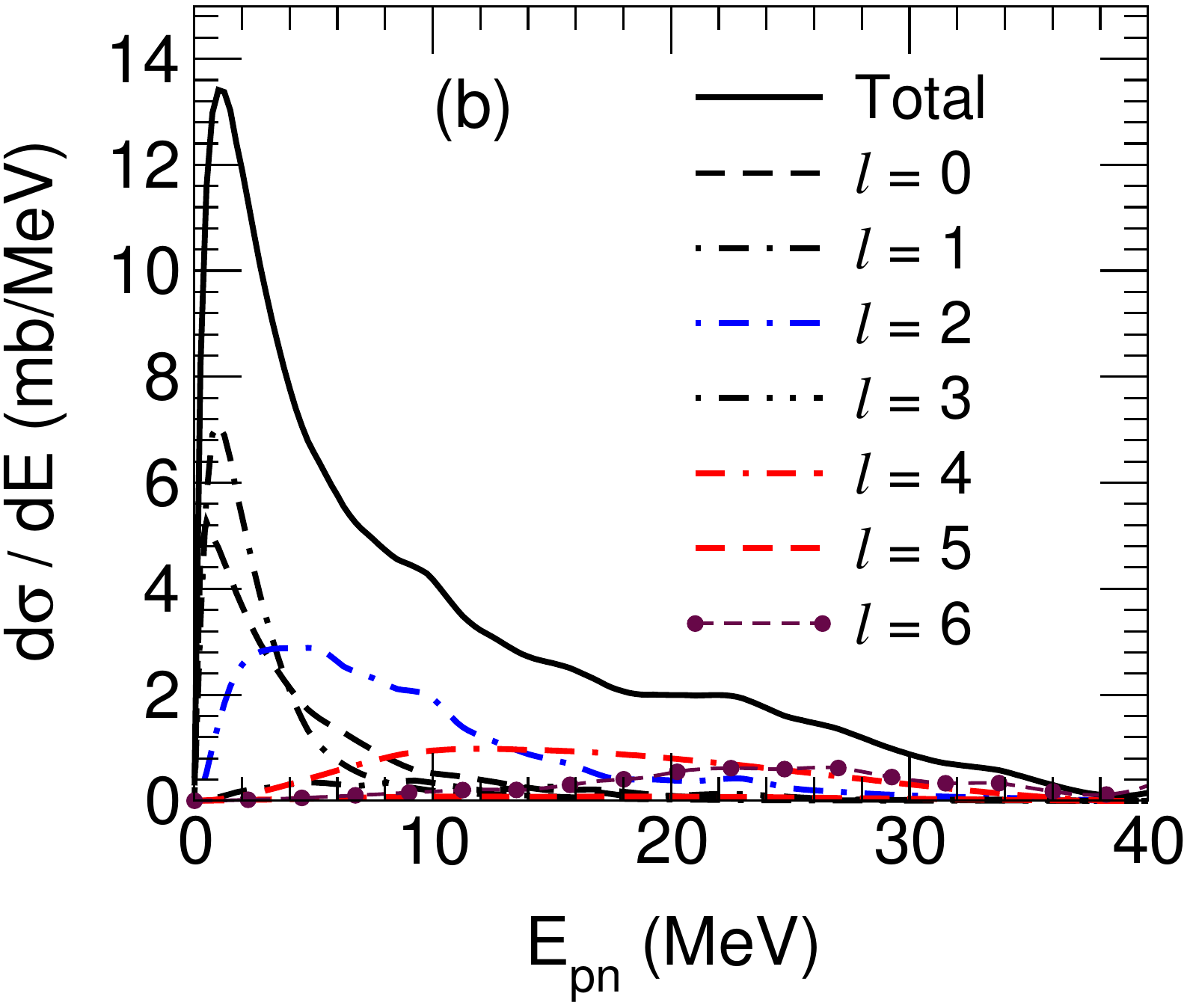}
\end{center}
\caption{\label{convergence}\small Comparing convergence of the breakup cross sections for (a) $^{12}$C(d,pn)$^{12}$C at $E_{\rm d}$ = 12 MeV and (b) $^{48}$Ca(d,pn)$^{48}$Ca at $E_{\rm d}$ = 56 MeV. Contributions of the different n-p partial waves to the energy distribution, plotted as a function of the n-p relative energy. }
\end{figure}

CDCC was shown to fail for breakup observables, for reactions performed at low deuteron energies \cite{upadhyay12}. While CDCC makes a choice of one Jacobi basis set for the expansion of the three-body wave-function, namely that of the deuteron, inspection of the Faddeev calculations demonstrated the importance of the proton and neutron components. Encapsulating these components within the CDCC basis, resulted in very slow convergence. In Fig.\ref{convergence} we show the contribution of each $np$ partial wave ($l$) to the breakup energy distribution for the reaction on $^{12}$C at 12 MeV (fig. a) and for the reaction on $^{48}$Ca at 56 MeV (fig. b). For $^{48}$Ca, the use of $l_{max}=6$ was sufficient to obtain convergence, as shown in the comparison with Faddeev \cite{upadhyay12}. For $^{12}$C, the predictions using $l_{max}=6$ were very far from the results obtained with Faddeev. We should  point out that for this particular reaction, calculations with $l_{max}=8$ become numerically unfeasible with the numerical methods presently used to solve the coupled differential equations in {\sc fresco}.

Moreover, if one estimated an error due to lack of convergence, based on the difference between the CDCC cross section with $l_{max}=4$ and $l_{max}=6$, as is customary, one would be mislead to believe in a much smaller 
error than the real error.  Even though Fig.\ref{convergence} appears to suggest that partial waves above $l_{max}=6$ will not contribute very much individually, the contributions do add up at large $np$ relative energies, and produce a very different shape for the distribution (see \cite{upadhyay12}. Ultimately, the rate of convergence is too slow.

\section{Outlook}
\vspace{0.5cm}

In this work we discuss some important developments in reaction theories, based on a three-body formulation, used to analyse deuteron induced reactions. We summarize the results obtained comparing the CDCC method,
and its more practical approximation ADWA, with the solution of the Faddeev equations in momentum-space.
Presently there is a better understanding of the validity of approximations in ADWA and CDCC when computing breakup and transfer cross sections. The fact that ADWA and CDCC are not good approximations for transfer at intermediate to high energy already calls for an alternative formulation. Most critically, at present no  benchmarks exist for heavier systems. Faddeev calculations for (d,p) on heavy nuclei are not possible due to the difficulty of handling the Coulomb interaction. It has thus become clear to the TORUS collaboration that indeed a new formalism \cite{akram12}, more efficient at handling the three-body equations, which provides stable solutions for a wide variety of cases, including  a wide range of relevant beam energies (5-50 MeV/u), light and heavy systems, well bound and loosely bound nuclei, is absolutely necessary in the field.

When the input for the reaction theory is well constrained by either data or the underlying microscopic structure, and an accurate solution to the three-body scattering problem is possible for the wide range of interesting cases to be explored, the full scientific potential of the existing and upcoming new generation facilities will be realized.

\vspace{1cm}
We thank all members of the TORUS collaboration for useful discussions.
The work of N.J.U. was supported by the Department of Energy topical collaboration grant DE-SC0004087.
The work of  F.M.N. was partially supported by the National Science Foundation grant PHY-0800026 and the Department of Energy through grant DE-FG52-08NA28552.

\end{document}